\documentclass[10pt]{article}

\usepackage[english]{babel}

\usepackage{graphics,color}
\usepackage{epsfig}

\begin{document}
\date{}
\begin{center}
{\Large\bf Sudden Death of Entanglement induced by a minimal thermal environment}
\end{center}
\begin{center}
{\normalsize G.L. De\c cordi and A. Vidiella-Barranco \footnote{vidiella@ifi.unicamp.br}}
\end{center}
\begin{center}
{\normalsize{ Gleb Wataghin Institute of Physics, University of Campinas - UNICAMP}}\\
{\normalsize{ 13083-859,   Campinas,  SP,  Brazil}}\\
\end{center}
\begin{abstract}
We study the dynamics of two interacting two-level systems (qubits) having one of them isolated and the other 
coupled to a single mode electromagnetic field in a thermal state. The field plays the role of a small environment, 
in contrast to the usual approach of modeling an environment via a thermal reservoir with many degrees of freedom.
We find the analytical solution of the proposed model, which allows us to investigate the consequences of the coupling 
to the small environment on characteristic quantum features of the two-qubit system. We study the time 
evolution of quantum entanglement and coherence, verifying the dependence on the relevant coupling constants
as well as the influence of the effective temperature of the environment. Interestingly, we find that both sudden 
death and sudden birth of entanglement may occur in such a simple system. We also discuss a different partition, 
in which the isolated qubit is considered to be coupled to a composite environment, constituted by the other 
qubit plus the field mode.
\end{abstract}
\section{Introduction}
A system constituted by two interacting qubits is one of the simplest quantum systems displaying joint nonclassical 
features such as entanglement. It is also well know that the coupling to a surrounding environment may affect such 
properties in a destructive way, e.g., causing decoherence \cite{zeh96}. Environments are normally modeled by a thermal 
reservoir, namely, a large number of quantum subsystems that are assumed to be coupled to the system of interest. In recent 
years, there has been renewed interest in the issue of derivation of appropriate master equations describing the dynamics of 
quantum coupled systems interacting with thermal baths 
\cite{kosloff14,johnson15,volovich16,avb17a,adesso17,brunner17,cresser17,plenio18,antezza18,zhang18a,zhang18b,zambrini19}. 
A matter of concern is that, for a composite quantum system and a given system-bath interaction Hamiltonian, the 
corresponding master equation may be not unique. As a consequence, it is likely to obtain different results regarding 
the evolution of the system, as the master equation may be derived by taking into account (or not) the interaction 
among the subsystems. If such an interaction is somehow included, we have the so-called {\it global} 
({\it microscopic}) model; 
otherwise the model is named {\it local} ({\it phenomenological}). Indeed, as it was already found in early works 
\cite{walls70,walls73,shibata79}, discrepancies could arise in the case of having strongly coupled subsystems interacting with
thermal baths. Nevertheless, as it has been recently pointed out, differences may also occur in the weak coupling regime 
\cite{avb17a,cresser92}, particularly if the bath is at finite temperature. It is worth noting that the  
additivity assumption of Lindblad terms in master equations has also been questioned in both the {\it local} approach as 
well as in the {\it global} approach, as discussed in \cite{plenio18}. 
Nevertheless, we have witnessed a predominant use of {\it local} master equations without much discussion 
about their validity. 
Yet, given the recent progress towards quantum technologies \cite{kim17,avb17b}, it is of importance an accurate 
description of the environmental influences on quantum systems, which means that one has to be sure that appropriate 
master equations are being employed. We thus believe there is need of further work on this particular theme.
At the same time, we have been witnessing an increasing control 
over elementary quantum systems which may be subjected to environments having a small number of degrees of freedom
(small environments). In fact, even a single uncontrollable subsystem coupled to the system of interest may cause a 
considerable disturbance. As examples of possible minimal environments, we may cite spins \cite{lombardi10,mirkin20}, a single two 
level system \cite{avb03,avb14,avb18} and one or two modes of the quantized electromagnetic field  
\cite{avb05,hanggi09,ashhab14,avb16}. In summary, we may be facing multiple challenges regarding the study of 
coupled quantum systems immersed in environments that might also be of different sizes, and in 
which the perturbative methods normally employed, e.g., {\it local approaches}, may not be applicable. Thus, simple
models could be useful to achieve a deeper understanding of the behaviour of such elementary systems.

In this contribution we are going to discuss a basic model which is related to the two-qubit system in interaction
with a thermal bath already studied in reference \cite{avb17a}. Consider two coupled two-level systems (qubits), being one 
isolated (qubit 1) and the other (qubit 2) interacting with a single mode thermal electromagnetic field, which will be 
playing the role of an (minimum) environment in place of a (multimode) thermal bath. We will investigate the dynamics 
and quantum properties of the two-qubit system for different effective temperatures of the field, as well as different 
system-environment coupling strengths ($g$). The model considered here admits analytical solution under the rotating wave 
approximation, and we may explore a wider range of values of $g$, in contrast to the inherent limitations of the 
(Markovian) master equations \cite{avb17a}, valid only in the weak (system-environment) coupling regime. Even though it is 
not our aim to mimic the action of a large reservoir using such a simple model, we could still make comparisons between some 
typical results from master equation-based methods and our simplified model. Yet, we would like to verify to what extent a 
small environment of that kind is able to degrade the quantum properties of the two-qubit system. 
In order to accomplish that, we are going to investigate the evolution of quantities that characterize nonclassical 
behaviour of the two-qubit system, such as entanglement and quantum coherence. We will also discuss the evolution of the 
state purity (linear entropy) of qubit 1 as if it were coupled to a ``small composite environment", constituted by qubit 2 
and the field mode. Unfortunately, quantum entanglement, a fundamental non-classical phenomenon and an important resource 
for quantum technologies, is quite fragile in the presence of external environments. 
For instance, considering a two-qubit system, entanglement may suddendly disappear in a finite time due to spontaneous 
emission, a process known as ``entanglement sudden death" (ESD) \cite{horodecki01,eberly04,eberly09}. 
Several interesting aspects of ESD have been uncovered in the past years. For example, while entanglement may 
remain null for a time interval during the evolution of a quantum system, it may suddendly revive, 
which is known as ``entanglement sudden birth" (ESB) \cite{tanas08,lopez08}. In \cite{lopez08} the authors 
consider a single mode cavity coupled to a multimode reservoir, in such a way that the cavity-reservoir system behaves,
for specific initial conditions, as an effective two-qubit system. 
They study how entanglement is distributed in a system
of two-cavities initially entangled (one photon each) plus two reservoirs in their vacuum states and demonstrate the 
ocurrence of ESB. Sudden-changes in multipartite entanglement are also investigated in \cite{rau14}. 
Both ESD and ESB have already been experimentally verified \cite{davidovich07,aguilar14}, and general conditions for 
the appearence of ESD in two-qubit systems coupled to thermal reservoirs 
have been discussed \cite{cunha07,napolitano11}. We remark that ESD is normally associated to a damping processes, 
that may be dissipative or not (e.g., non-dissipative dephasing noise) \cite{eberly04,lopez08,rau14,eberly06a}. 
Curiously, ESD also occurs in a (non-interacting) two-qubit system having each qubit coupled solely to a single mode 
field within a lossless cavity \cite{eberly06b}, i.e., without being in contact with a large number of sub-systems 
(multimode thermal reservoir). Naturally, in that case, as discussed in \cite{eberly06b}, the typical time intervals 
in which the system is unentangled are relatively short, and the entanglement births are periodic.  
If we now consider two completely isolated coupled qubits (absence of an environment), an entanglement quantifier e.g., 
the Concurrence relative to the two-qubit state will be simply a periodic function of time. Of course disturbances to 
such a regular evolution are expected if the system is interacting with an external reservoir. 
As it is reported in \cite{avb17a}, in the case of a multimode thermal bath, ESD may occur during the evolution, 
even though the steady state of the two-qubit system might exhibit some amount of entanglement.
Here our purpose is to provide novel insights about the dynamics of quantum coupled systems subjected to an environment 
via a simple model which allows analytical solution. Our model of two coupled qubits in contact with a single system 
reservoir which can be in a highly mixed state is in contrast to previous studies that typically 
consider two uncoupled subsystems (normally prepared in an entangled state) under the influence
of either multimode \cite{eberly04,lopez08} or single mode \cite{eberly06b} environments.
A preliminary study \cite{avb19} showed that ESD may also occur in our simple two (coupled) qubit model.
Yet, despite of all studies about ESD done in the past years, we believe this subject deserves further 
investigation, taking into account different scenarios, as the emergence of ESD during the evolution of coupled 
quantum systems may have a significant impact on the development of quantum technologies.

Our paper is organized as follows: in Section (2) we obtain the analytical solution of the model.  
In Section (3) we investigate the influence of the thermal field on the evolution of the qubit-qubit entanglement. 
In Section (4) we discuss the evolution of quantum coherence of the two-qubit state as well as the evolution of the 
isolated qubit, considering a ``composite reservoir" constituted by qubit 2 and the thermal field. We present our conclusions in
Section (5).

\section{The model}

The system we are going to discuss here consists of two 2-level atoms (qubits) plus a mode of the quantized field. 
The qubits are themselves coupled (via dipole-dipole interaction), while only one of them (qubit 2) is coupled to a 
single mode field (via Jaynes-Cummings interaction).
We assume equal qubit's transition frequencies $(\omega_{1}=\omega_{2})$ as well as resonance between qubit 2 and the
field $(\omega_{2}=\omega)$. The Hamiltonian $H$ of the system may be written as (in units of $\hbar$) $H=H_{0}+H_{1}$, with
\begin{equation}
H_{0}=\frac{\omega}{2}\sigma_{1z}+\frac{\omega}{2}\sigma_{2z}+\omega a^{\dagger}a,\label{eq:H0}
\end{equation}
and
\begin{equation}
H_{1}=\lambda\left(\sigma_{1}^{+}\sigma_{2}^{-}+\sigma_{1}^{-}\sigma_{2}^{+}\right)+g\left(a\sigma_{2}^{+}+a^{\dagger}\sigma_{2}^{-}\right)\,.
\label{eq:H1}
\end{equation}

Here  
$\sigma_{iz}=\left|e_{i}\right\rangle \left\langle e_{i}\right|-\left|g_{i}\right\rangle \left\langle g_{i}\right|$,
$\sigma_{i}^{+}=\left|e_{i}\right\rangle \left\langle g_{i}\right|$, and
$\sigma_{i}^{-}=\left|g_{i}\right\rangle \left\langle e_{i}\right|$ are the operator relative to the qubits ($i=1,2$), and
$a$ and $a^{\dagger}$ are the usual annihilation and creation field operators. The effective coupling constant between the two 
qubits is denoted by $\lambda$ and $g$ is the coupling constant between qubit 2 and the field (small environment). 
In Appendix A one may find details of the solution for $\rho_I(t)$ [see Eq. (\ref{eq:ro global inter})], the time-evolved joint 
density operator in the interaction representation, 
\begin{equation}
\rho_{I}\left(t\right)=e^{-i\,H_{1}t}\rho\left(0\right)e^{i\,H_{1}t},
\end{equation}
with an initial separable density operator given by
$\rho\left(0\right)  =  \rho_{q1}\left(0\right)\otimes\rho_{q2}\left(0\right)\otimes\rho_{f}\left(0\right)$, or

\begin{equation}
\rho\left(0\right)  =  \left|e_{1}\right\rangle \left\langle e_{1}\right|\otimes\left|g_{2}\right\rangle \left\langle g_{2}\right|
\otimes\sum_{n=0}^{\infty}P_n\left|n\right\rangle \left\langle n\right|\equiv \sum_{n=0}^{\infty}P_n
\left|e_{1},\,g_{2},\,n\right\rangle \left\langle e_{1},\,g_{2},\,n\right|.\label{eq:ro zero}
\end{equation}
Here
\begin{equation}
P_n=\frac{\overline{n}^{n}}{\left(1+\overline{n}\right)^{n+1}}\label{eq:dist termico}
\end{equation}
is the photon number distribution of the single mode thermal field state (frequency $\omega$) 
with mean photon number $\overline{n} = [\exp(\omega/k_B T)-1]^{-1}$ (effective temperature $T$).
The time evolved two-qubit density operator may be calculated by tracing over the field variables, i.e.,
$\rho_{q1,q2}\left(t\right)=Tr_{f}\left[\rho_I\left(t\right)\right]$, as shown in Appendix A 
[see Eq.(\ref{eq:ro atom termico App})]. 

\section{Influence of a small thermal environment: two-qubit entanglement}

The qubit-qubit interaction naturally leads to bipartite entanglement. One of the possible ways of quantifying the two-qubit
entanglement is via a function such as the Concurrence ${\cal C}$ \cite{wootters97}. In our model involving two qubits, we may compute 
${\cal C}(t)$, the Concurrence as a function of time, as follows 

\begin{equation}
\mathcal{C\mathrm{\left(t\right)}=\mathrm{max\,\left[0,\Lambda\left(t\right)\right]\,,}}
\end{equation}
with 

\begin{equation}
\Lambda\left(t\right)\equiv\sqrt{\xi_{1}\left(t\right)}-\sqrt{\xi_{2}\left(t\right)}-\sqrt{\xi_{3}\left(t\right)}-\sqrt{\xi_{4}\left(t\right)}\,.
\end{equation}
The quantities $\xi_{i}$ are the eigenvalues of the matrix $M\left(t\right)=\rho_{q1,q2}\left(t\right)\left(\sigma_{y}^{(1)}
\otimes\sigma_{y}^{(2)}\right)\rho^{*}_{q1,q2}\left(t\right)\left(\sigma_{y}^{(1)}\otimes\sigma_{y}^{(2)}\right)$
which should be placed in decreasing order, and $\sigma_{y}$ is Pauli's matrix. 

The two-qubit density operator $\rho_{q1,q2}\left(t\right)$ calculated in Appendix A
[see Equation(\ref{eq:ro atom termico App})] clearly belongs to the peculiar class of bipartite states 
called X states \cite{eberly07a}, i.e.,

\begin{equation}
\rho_{q1,q2}\left(t\right)=\left(\begin{array}{cccc}
\rho_{11} & 0 & 0 & 0\\
0 & \rho_{22} & \rho_{23} & 0\\
0 & \rho_{23}^{*} & \rho_{33} & 0\\
0 & 0 & 0 & \rho_{44}
\end{array}\right),
\end{equation}
where

\begin{equation}
\rho_{11}=\sum_{n=0}^{\infty}P_{n+1}\left|C_{1,\,n+1}\right|^{2}\,; \ \ \ \ \ \
\rho_{22}=\sum_{n=0}^{\infty}P_{n}\left|C_{2,\,n}\right|^{2},
\end{equation}

\begin{equation}
\rho_{23}=\sum_{n=0}^{\infty}P_{n}C_{2,\,n}C_{3,\,n}^{*}\,; \ \ \ \ \ \
\rho^*_{23}=\sum_{n=0}^{\infty}P_{n}C_{2,\,n}^{*}C_{3,\,n}
\end{equation}

\begin{equation}
\rho_{33}=\sum_{n=0}^{\infty}P_{n}\left|C_{3,\,n}\right|^{2}\,; \ \ \ \ \ \
\rho_{44}=\sum_{n=0}^{\infty}P_{n-1}\left|C_{4,\,n-1}\right|^{2}\,.
\end{equation}

Again, the expression for the coefficients $C_{j,n}$ may be found in Appendix A. 
The X states may arise in a variety of situations involving the quantum dynamics of two qubit systems,
normally when a partial trace is taken over a third subsystem (often a reservoir) 
\cite{avb05,eberly04,lopez08,rau14,vedral01,eberly07a}.
In particular, concerning such states, it is straightforward to obtain a compact expression for the 
Concurrence \cite{eberly07a}. In our case, the function $\Lambda\left(t\right)$ is given by

\begin{equation}
\Lambda\left(t\right)=2\left|\rho_{23}\right|-2\sqrt{\rho_{11}\rho_{44}}.\label{lambdafunction}
\end{equation}

The Concurrence ${\cal C}(t)$ is a periodic function of time for a completely 
isolated two-qubit system ($g = 0.0$), and thus the destructive action of an environment may be estimated
by observing the shortening of the amplitude of the oscillations during the evolution. 
Let us first consider qubit 2 weakly coupled to the field ($g = 0.1 \lambda$); we expect some disturbance even for 
relatively small excitations of the environment (e.g. $\overline{n} = 1$). In order to verify the effect of the 
environment, we have numerically evaluated the Concurrence, which is plotted as a function of time in 
Figure (1) for two values of the field mean excitation number: $\overline{n} = 1$ 
in Figure (1;a,b,c), and $\overline{n} = 10$ in Figure (1;d,e,f). In all plots we adopt the normalization $\lambda = 10.0$. 
We note in Figure (1;a,b,c) that the amplitude of the 
oscillations decrease steadily, although at later times occur periodic revivals. As one would expect, for larger values of 
$\overline{n}$, a stronger degradation of entanglement is verified: if $\overline{n} = 10$, the nearly 
periodic revivals are considerably more attenuated as shown in Figure (1;d,e,f), in contrast 
to the case in which $\overline{n} = 1$ [see Figure (1;a,b,c)]. An interesting aspect of this model is that the
phenomenon of ESD occurs during the evolution. This is already noticeable in the weaker coupling
case, i.e., the Concurrence becomes zero during short time intervals, as shown in Figure (1;d), for instance.
Naturally the ESD is expected to be more evident for a stronger system-environment coupling. If $g$ is increased to 
$g = 0.5\lambda$ (but still keeping $g$ smaller than $\lambda$), 
a typical pattern of rapid entanglement death and birth is observed, even for small excitation number, as we see in 
Figure (2;a,b,c). Yet, the time intervals having null entanglement are longer, and although there are sudden-births of 
entanglement during the evolution, the maximum values of the Concurrence are in average lower than those for $g = 0.1 \lambda$. 
Of course an even more destructive action of the thermal noise occurs if $\overline{n}$ is increased. As a matter of fact, 
entanglement is almost completely washed out, and no births are verified at longer time-scales if $\overline{n} = 10$ 
[see Figure (2;d,e,f)]. 

An alternative way of assessing the existence of ESD \cite{eberly07b} is via the function $\Lambda(t)$ defined above, 
a quantifier of the two-qubit state separability [see Eq. (\ref{lambdafunction})]. If $\Lambda$ becomes negative, 
the corresponding density operator $\rho_{q1,q2}$ describes a mixed separable state, and thus null entanglement is detected. 
We have plotted the function $\Lambda(t)$ as a function of time, for $g = 0.1 \lambda$ in Figure (3) and 
for $g = 0.5 \lambda$ in Figure (4). In particular, the negativity of the $\Lambda$ function makes evident that the 
suppression of entanglement may persist for long time intervals, as shown in Figure (4;f). Also, it is possible to 
estimate, at particular times, if the two-qubit state is close or not of becoming entangled, depending on the amount 
negativity in the $\Lambda$ function. 

\section{Influence of a small thermal environment: two-qubit quantum coherence and state purity of the isolated qubit}

\subsection{Quantum coherence}

Quantum coherence, a fundamental concept in quantum theory, has been the object of discussions in the recent literature 
\cite{plenio14,plenio17}. As a matter of fact, quantum coherence is a resource that may be quantified \cite{plenio14}, 
in analogy to what has been already done regarding quantum entanglement. A simple and direct measure of coherence is given 
by the quantifier $C_{l_1}$ ($l_1$ norm of coherence), defined as

\begin{equation}
C_{l_1} = \sum_{i,j,i\neq j} \left| \rho_{ij}\right| =
\left|\sum_{n=0}^{\infty}P_{n}\,C_{2,\,n}C_{3,\,n}^{*}\right|+\left|\sum_{n=0}^{\infty}P_{n}\,C_{2,\,n}^{*}C_{3,\,n}\right|,
\end{equation}
where $\rho_{ij} \equiv \langle i|\rho_{q1,q2}| j\rangle$ are the matrix elements of the two-qubit density operator. 
In what follows we are going to analyze the influence of the small thermal environment on the quantum coherence of 
the two-qubit system. 

Our expectation is that the thermal environment will degrade quantum coherence. Indeed, as shown in Figure (5),
we observe the transition from a perfectly periodic to a more irregular behaviour. For instance, for a weaker coupling 
($g = 0.1 \lambda$), the
time evolution of coherence is very similar to the evolution of entanglement (Concurrence), as shown in 
Figure (5;a,b,c). If the qubit 2-field coupling is increased ($g = 0.5 \lambda$), we observe an overall 
stronger degradation of quantum coherence, characterized by irregular (and damped) oscillations, as seen 
in Figure (6). Nevertheless, differently from what happens to quantum entanglement, 
the vanishing of quantum coherence in a finite time is not verified during the evolution.  

\subsection{Purity of the qubit 1: linear entropy}

Now we would like to discuss the influence of the environment on the (reduced) dynamics of qubit 1.  
From the system's density operator $\rho_I(t)$, we calculate the reduced density operator of qubit 1
by tracing over the field and qubit 2, i.e., $\rho_{q1}(t) = Tr_{q2,f}\left[\rho_I(t)\right]$.
Note that everything goes as if qubit 1 was coupled to a ``composite bath", constituted by the field and 
qubit 2. In order to characterize the purity of the qubit 1 state we use the linear entropy, defined as 
$S\left(t\right)=1-Tr_{q1}\left(\rho_{q1}^{2}\right)$. In this case the entropy may be written as $S\left(t\right)=1-\rho_{ee}^{2}-\rho_{gg}^{2}$, 
and after some algebra one is able to express $S(t)$ as a function of the inversion 
$W\left(t\right)=\rho_{ee}\left(t\right)-\rho_{gg}\left(t\right) = 1-2\rho_{gg}\left(t\right)$, or
\begin{equation}
	S\left(t\right) = \frac{1}{2}-\frac{W(t)^{2}}{2}\,.\label{eq:S+W}
\end{equation}
Using the matrix element $\rho_{gg}$ calculated in the Appendix, we obtain the inversion,

\begin{eqnarray}
	W\left(t\right) &=& 1-2\sum_{n=0}^{\infty}P_n\left[\left|A_{32}^{\left(n\right)}\right|^{2}
	+\left|A_{42}^{\left(n\right)}\right|^{2}\right]\nonumber \\
	& = & 1-\frac{2}{k^{2}}\sum_{n=0}^{\infty}\frac{P_{n}}{\beta_{n}^{2}}\left\{ \frac{1+\left(4n+3\right)k^{2}}{2k^{2}}+\left[\frac{\left(1-\beta_{n}\right)k^{2}-1}{4k^{2}}\right]\cos(2\omega_{+,n}t)+
	\left[\frac{\left(1+\beta_{n}\right)k^{2}-1}{4k^{2}}\right]\cos(2\omega_{-,n}t)-\right.\nonumber \\
	&  & \left.\left[n+1-\sqrt{n\left(n+1\right)}\right]\cos\left[\left(\omega_{+,n}+\omega_{-,n}\right)t\right]-
	\left[n+1+\sqrt{n\left(n+1\right)}\right]\cos\left[\left(\omega_{+,n}-\omega_{-,n}\right)t\right]\right\}.\label{inversion}
\end{eqnarray}

The expression for the quantities ($\omega_{\pm}$ and $\beta_n$) in Equation (\ref{inversion}) above
may be found in the Appendix. We remark that the parameter $k$ is the 
ratio between the qubit 2-field (environment) coupling and the qubit 1-qubit 2 coupling, i.e., $k \equiv g/\lambda$.

The purity of qubit 1 in the absence of an environment is a periodic function of time, as initially independent qubits
periodically get entangled and disentangled as time goes on.
We have numerically evaluated the linear entropy of qubit 1, $S(t)$, which is plotted as a function of time in 
Figure (7), for two values of the field mean 
excitation number: $\overline{n} = 1$ in Figure (7;a,b,c), and $\overline{n} = 10$ in Figure (7;d,e,f). 
Note that as we increase $\overline{n}$, there is a higher degradation of qubit 1's purity, as expected. 
Besides, for larger $\overline{n}$ the quasi-revivals are also less defined (within the time window considered). 

We also analyze the situation of stronger coupling to the field, e.g., for 
$g = 0.5 \lambda$, which is shown in Figure (8). We note a different
pattern compared to the former case (weaker coupling): besides revealing a more irregular evolution, it is clear
the drifting of qubit 1 towards a less pure state, specially for larger $\overline{n}$ [see Figure (8;d,e,f)]. We
would like to point out a peculiar behaviour of the entropy in this case. We note that the slope of $S(t)$ (for short times) 
is less steep for $\overline{n} = 10$ [Figure (8d)] than for $\overline{n} = 1$ [Figure (8a)]. 
This happens because in the beginning of the evolution the interaction between qubits leads to qubit-qubit entanglement 
(and consequently to mixedness). Yet, the noise from the small environment gradually comes into play, taking the state 
of qubit 2 to a statistical mixture of ground and excited states. This momentarily reduces the capability of entanglement 
between the sub-systems, as one of the component states in the mixture is an eigenstate of the Hamiltonian. However, 
as time goes on, the action of the noisy system (field) prevails, progressively leading qubit 1 to a less and less pure state.

\section{Conclusions}

We have studied a simple toy-model of a composite (two-qubit) system in interaction with a very small environment, 
namely a single mode thermal field. We wanted to assess how damaging the action of a small non-dissipative environment 
can be to the quantum properties of the two-qubit system, in contrast to the usually adopted master-equation-based approach, 
in which the system is assumed to be coupled to a large dissipative environment, e.g., a multimode thermal 
bath \cite{avb17a}. Despite the smallness of the environment we are here considering, due to the spread in excitation 
number of the thermal state, the resulting incommensurate frequencies may give rise to an irreversible-like behaviour. One 
advantage of a simplified approach is that the model admits analytical solution, and there is no need of further 
approximations such as perturbation theory, for instance. Thus, we may ``tune" the coupling constants and have a broader 
picture of the environment's impact on the dynamics of the system. Needless to say that maximally entangled states 
periodically arise in the two-qubit system if the coupled qubits are isolated. Interestingly, we have found that the 
interaction with a minimal environment can give rise to the ``sudden-death of entanglement" phenomenon 
\cite{horodecki01,eberly04,eberly09}. Entanglement births \cite{tanas08,lopez08} also take place, and the length of the time intervals 
having null entanglement strongly depends on the couplings as well as on the mean photon number of the thermal environment. 
Other nonclassical features, viz., the state purity of qubit 1 and the quantum coherence of the two-qubit 
system, may also be considerably degraded, even though an effect analogous to a sudden death does not take place 
in those cases. 

\section*{Acknowledgements}

This work was also supported by CNPq 
(Conselho Nacional para o Desenvolvimento Cient\'\i fico e Tecnol\'ogico), 
through the INCT-IQ (National Institute for Science and Technology of Quantum Information), 
grant  N${\textsuperscript{\underline{o}}}$ 465469/2014-0, and FAPESP 
(Funda\c c\~ao de Amparo \`a Pesquisa do Estado de S\~ao Paulo),
grant N${\textsuperscript{\underline{o}}}$ 2008/57856-6, Brazil.

\appendix

\section{Solution of the model: two coupled qubits interacting with a single mode field}

We may expand the system's state vector (interaction picture) as:
\begin{equation}
\left|\psi_{I}(t)\right\rangle =C_{1,n}\left(t\right)\left|e_{1},e_{2},n-1\right\rangle +C_{2,n}\left(t\right)\left|e_{1},g_{2},n\right\rangle +C_{3,n}\left(t\right)\left|g_{1},e_{2},n\right\rangle +C_{4,n}\left(t\right)\left|g_{1},g_{2},n+1\right\rangle \,.\label{eq:sol tentativa}
\end{equation}
After substituting the ansatz above in the Schr\"odinger equation, we obtain the following system of coupled differential equations
for the coefficients $C_j$,

\begin{eqnarray}
i\,\dot{C}_{1,n} & = & g\sqrt{n}\,C_{2,n}\nonumber \\
i\,\dot{C}_{2,n} & = & g\sqrt{n}\,C_{1,n}+\lambda\,C_{3,n}\nonumber \\
\label{eq:coupled}\\
i\,\dot{C}_{3,n} & = & \lambda\,C_{2,n}+g\sqrt{n+1}\,C_{4,n}\nonumber \\
i\,\dot{C}_{4,n} & = & g\sqrt{n+1}\,C_{3,n}\,.\nonumber 
\end{eqnarray}

The solution of the system (\ref{eq:coupled}) may be expressed as
\begin{equation}
C_{j,n}\left(t\right)=\sum_{m=1}^{4}A_{jm}^{\left(n\right)}\left(t\right)c_{m}\left(0\right)\,,\label{eq:comb l}
\end{equation}
with $j=1,2,3,4$ and $A_{jm}^{\left(n\right)}=A_{mj}^{\left(n\right)}$.
The $\left(n\right)$ in the matrix elements refers to the dimensionality of the 
Fock state basis, $n=0,1,2,\ldots$ 

The relevant matrix elements in Equation (\ref{eq:comb l}) are given by
\begin{eqnarray}
A_{11}^{\left(n\right)}\left(t\right) & = & \frac{1}{r_{n}}\left[\left(\omega_{+,n}^{2}-b_{n}^{2}-\lambda^{2}\right)\cos(\omega_{+,n}t)-\left(\omega_{-,n}^{2}-b_{n}^{2}
-\lambda^{2}\right)\cos(\omega_{-,n}t)\right]\nonumber \\
A_{12}^{\left(n\right)}\left(t\right) & = & \frac{i\,a_{n}}{r_{n}}\left[\frac{\left(b_{n}^{2}-\omega_{+,n}^{2}\right)}{\omega_{+,n}}\sin(\omega_{+,n}t)-\frac{\left(b_{n}^{2}
	-\omega_{-,n}^{2}\right)}{\omega_{-,n}}\sin(\omega_{-,n}t)\right]\nonumber \\
A_{13}^{\left(n\right)}\left(t\right) & = & \frac{\lambda\,a_{n}}{r_{n}}\left[\cos(\omega_{+,n}t)-\cos(\omega_{-,n}t)\right]\nonumber \\
A_{14}^{\left(n\right)}\left(t\right) & = & -\frac{i\,\lambda a_{n}b_{n}}{r_{n}}\left[\frac{\sin(\omega_{+,n}t)}{\omega_{+,n}}-\frac{\sin(\omega_{-,n}t)}{\omega_{-,n}}\right]\nonumber \\
A_{22}^{\left(n\right)}\left(t\right) & = & \frac{1}{r_{n}}\left[\left(\omega_{+,n}^{2}-b_{n}^{2}\right)\cos(\omega_{+,n}t)-\left(\omega_{-,n}^{2}-b_{n}^{2}\right)\cos(\omega_{-,n}t)\right]\nonumber \\
\label{eq:sol}\\
A_{23}^{\left(n\right)}\left(t\right) & = & -\frac{i\,\lambda}{r_{n}}\left[\omega_{+,n}\sin(\omega_{+,n}t)-\omega_{-,n}\sin(\omega_{-,n}t)\right]\nonumber \\
A_{24}^{\left(n\right)}\left(t\right) & = & \frac{\lambda\,b_{n}}{r_{n}}\left[\cos(\omega_{+,n}t)-\cos(\omega_{-,n}t)\right]\nonumber \\
A_{33}^{\left(n\right)}\left(t\right) & = & \frac{1}{r_{n}}\left[\left(\omega_{+,n}^{2}-a_{n}^{2}\right)\cos(\omega_{+,n}t)-\left(\omega_{-,n}^{2}-a_{n}^{2}\right)\cos(\omega_{-,n}t)\right]\nonumber \\
A_{34}^{\left(n\right)}\left(t\right) & = & \frac{i\,b_{n}}{r_{n}}\left[\frac{\left(a_{n}^{2}-\omega_{+,n}^{2}\right)}{\omega_{+,n}}\sin(\omega_{+,n}t)-\frac{\left(a_{n}^{2}
	-\omega_{-,n}^{2}\right)}{\omega_{-,n}}\sin(\omega_{-,n}t)\right]\nonumber \\
A_{44}^{\left(n\right)}\left(t\right) & = & \frac{1}{r_{n}}\left[\left(\omega_{+,n}^{2}-a_{n}^{2}-\lambda^{2}\right)\cos(\omega_{+,n}t)-\left(\omega_{-,n}^{2}-a_{n}^{2}-\lambda^{2}\right)\cos(\omega_{-,n}t)\right]\,,\nonumber 
\end{eqnarray}
with
\begin{equation}
a_{n}=g\sqrt{n}\,,\qquad b_{n}=g\sqrt{n+1}\,,\label{eq:an+bn}
\end{equation}

\begin{equation}
r_{n}=\sqrt{\left(g^{2}+\lambda^{2}\right)^{2}+4ng^{2}\lambda^{2}}\,,\label{eq:rn}
\end{equation}

\begin{equation}
\omega_{\pm,n}=\frac{1}{\sqrt{2}}\sqrt{\left(2n+1\right)g^{2}+\lambda^{2}\pm\sqrt{\left(g^{2}+\lambda^{2}\right)^{2}+4ng^{2}\lambda^{2}}}\,.\label{eq:omg}
\end{equation}

We may define the ratio of the qubit 1-qubit 2 coupling to the qubit 2-field as $k\equiv g/\lambda$,
and rewrite (\ref{eq:rn}) and (\ref{eq:omg}) in terms of $k$
\begin{equation}
r_{n}=\lambda^2\,\sqrt{\left(1+k^{2}\right)^{2}+4\,nk^{2}}\,,\label{eq:rn+k}
\end{equation}

\begin{equation}
\omega_{\pm,n}=\frac{\lambda}{\sqrt{2}}\sqrt{1+\left(2n+1\right)k^{2}\pm\sqrt{\left(1+k^{2}\right)^{2}+4\,nk^{2}}}\,.\label{eq:omg+k}
\end{equation}

If we also define the auxiliary parameters
\begin{equation}
\alpha_{n}=1+\left(2n+1\right)k^{2}\,,\label{eq:def alfa}
\end{equation}
and
\begin{equation}
\beta_{n}=\sqrt{\left(1+k^{2}\right)^{2}+4\,nk^{2}}\,,\label{eq:def beta}
\end{equation}
the characteristic frequencies may be expressed as
\begin{equation}
\omega_{\pm,n}=\frac{\lambda}{\sqrt{2}}\sqrt{\alpha_{n}\pm\beta_{n}}\,.\label{eq:omg+alfa+beta}
\end{equation}

For the specific initial condition $|\psi(0)\rangle = |e_1,g_2,n\rangle$, we have that 
$c_{1}\left(0\right)=0$, $c_{2}\left(0\right)=1$,
$c_{3}\left(0\right)=0$ and $c_{4}\left(0\right)=0$, so that the amplitudes $C_{j,n}$ become
\begin{equation}
C_{1,n}\left(t\right)=A_{12}^{\left(n\right)}\left(t\right)\,,\quad C_{2,n}\left(t\right)=A_{22}^{\left(n\right)}\left(t\right)\,,\quad 
C_{3,n}\left(t\right)=A_{23}^{\left(n\right)}\left(t\right)\,,\quad C_{4,n}\left(t\right)=A_{24}^{\left(n\right)}\left(t\right)\,.\label{eq:amp prob Fock}
\end{equation}

Thus the system's joint density operator may be expressed as:

\begin{eqnarray}
\rho_I\left(t\right)  =  \sum_{n=0}^{\infty}P_n\left[C_{1,n}\left(t\right)\left|e_{1},e_{2},n-1\right\rangle 
+C_{2,n}\left(t\right)\left|e_{1},g_{2},n\right\rangle +C_{3,n}\left(t\right)\left|g_{1},e_{2},n\right\rangle 
+C_{4,n}\left(t\right)\left|g_{1},g_{2},n+1\right\rangle \right]\times\nonumber \\
 \left[C_{1,n}^*\left(t\right)\left\langle e_{1},e_{2},n-1\right|
+C_{2,n}^*\left(t\right)\left\langle e_{1},g_{2},n\right|+C_{3,n}^*\left(t\right)\left\langle g_{1},e_{2},n\right|
+C_{4,n}^*\left(t\right)\left\langle g_{1},g_{2},n+1\right|\right].\label{eq:ro global inter}
\end{eqnarray}

After tracing over the field, we obtain the following two-qubit system density operator

\begin{equation}
\rho_{q1,q2}\left(t\right)=Tr_{f}\left[\rho_I\left(t\right)\right]\,,\label{eq:def ro atom}
\end{equation}
\begin{eqnarray}
\rho_{q1,q2}\left(t\right)  =  
 \sum_{n=0}^{\infty}P_{n+1}\left|C_{1,\,n+1}\right|^{2}\left|e_{1},\:e_{2}\right\rangle \left\langle e_{1},\:e_{2}\right|
+\sum_{n=0}^{\infty}P_n\left|C_{2,\,n}\right|^{2}\left|e_{1},\:g_{2}\right\rangle \left\langle e_{1},\:g_{2}\right|+\nonumber \\
 \sum_{n=0}^{\infty}P_n\left|C_{3,\,n}\right|^{2}\left|g_{1},\:e_{2}\right\rangle \left\langle g_{1},\:e_{2}\right|
+\sum_{n=0}^{\infty}P_{n-1}\left|C_{4,\,n-1}\right|^{2}\left|g_{1},\:g_{2}\right\rangle \left\langle g_{1},\:g_{2}\right|+\nonumber \\
\sum_{n=0}^{\infty}P_n\,C_{2,\,n}C_{3,\,n}^*\left|e_{1},\:g_{2}\right\rangle \left\langle 
g_{1},\:e_{2}\right|+\sum_{n=0}^{\infty}P_n\,C_{2,\,n}^*C_{3,\,n}\left|g_{1},\:e_{2}\right\rangle \left\langle e_{1},\:g_{2}\right| \,.\label{eq:ro atom termico App}
\end{eqnarray}

The density operator for qubit 1 is obtained after tracing over qubit 2, $\rho_{q1}\left(t\right)=Tr_{q2}\left[\rho_{q1,q2}\left(t\right)\right]$,
resulting in
\begin{equation}
\rho_{q1}\left(t\right)=\rho_{ee}\left(t\right)\left|e_{1}\right\rangle \left\langle e_{1}\right|
+\rho_{gg}\left(t\right)\left|g_{1}\right\rangle \left\langle g_{1}\right|\,,\label{eq:ro a1 termico}
\end{equation}
where
\begin{equation}
\rho_{ee}\left(t\right)  =  \sum_{n=0}^{\infty}P_{n+1}\left|C_{1,\,n+1}\right|^{2}
+\sum_{n=0}^{\infty}P_n\left|C_{2,\,n}\right|^{2}\,,\label{eq:roee term}
\end{equation}
and
\begin{equation}
\rho_{gg}\left(t\right)  =  \sum_{n=0}^{\infty}P_n\left|C_{3,\,n}\right|^{2}
+\sum_{n=0}^{\infty}P_{n-1}\left|C_{4,\,n-1}\right|^{2}.\label{eq:rogg term}
\end{equation}



\newpage

\begin{figure}
	\centering
	\includegraphics[scale=0.8]{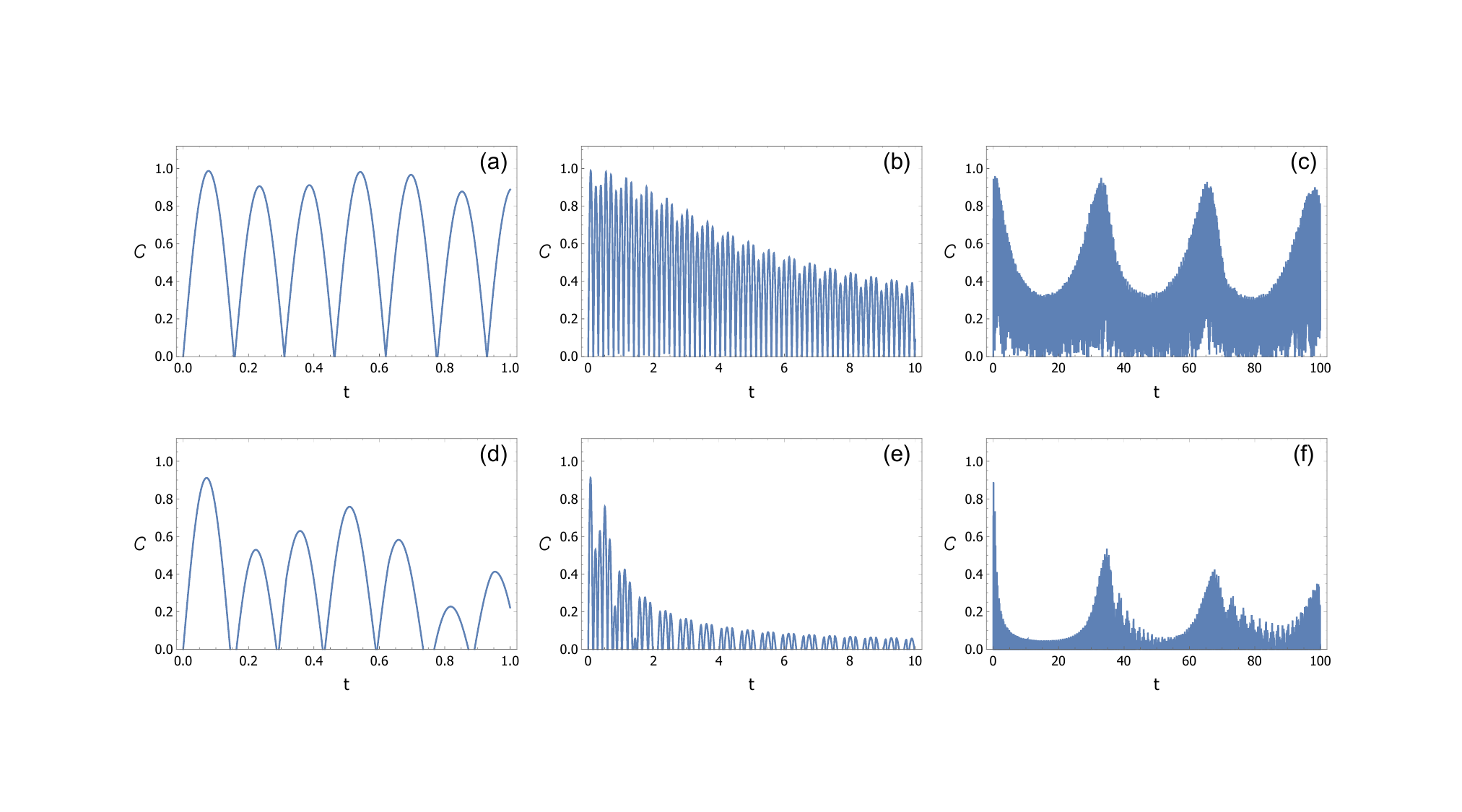}
	\caption{Concurrence of the two-qubit system as a function of time for $g = 0.1 \lambda$ 
		and for different time-scales. The mean photon number of the field is 
		$\overline{n} = 1$ in (a), (b) and (c), and $\overline{n} = 10$ in (d), (e) and (f).
		The initial state is $\rho\left(0\right)  =  
		\left|e_{1}\right\rangle \left\langle e_{1}\right|\otimes\left|g_{2}\right\rangle \left\langle g_{2}\right|
		\otimes\sum_{n=0}^{\infty}P_n\left|n\right\rangle \left\langle n\right|$.} 
	\label{figure1}	
\end{figure} 

\begin{figure}
	\centering
	\includegraphics[scale=0.8]{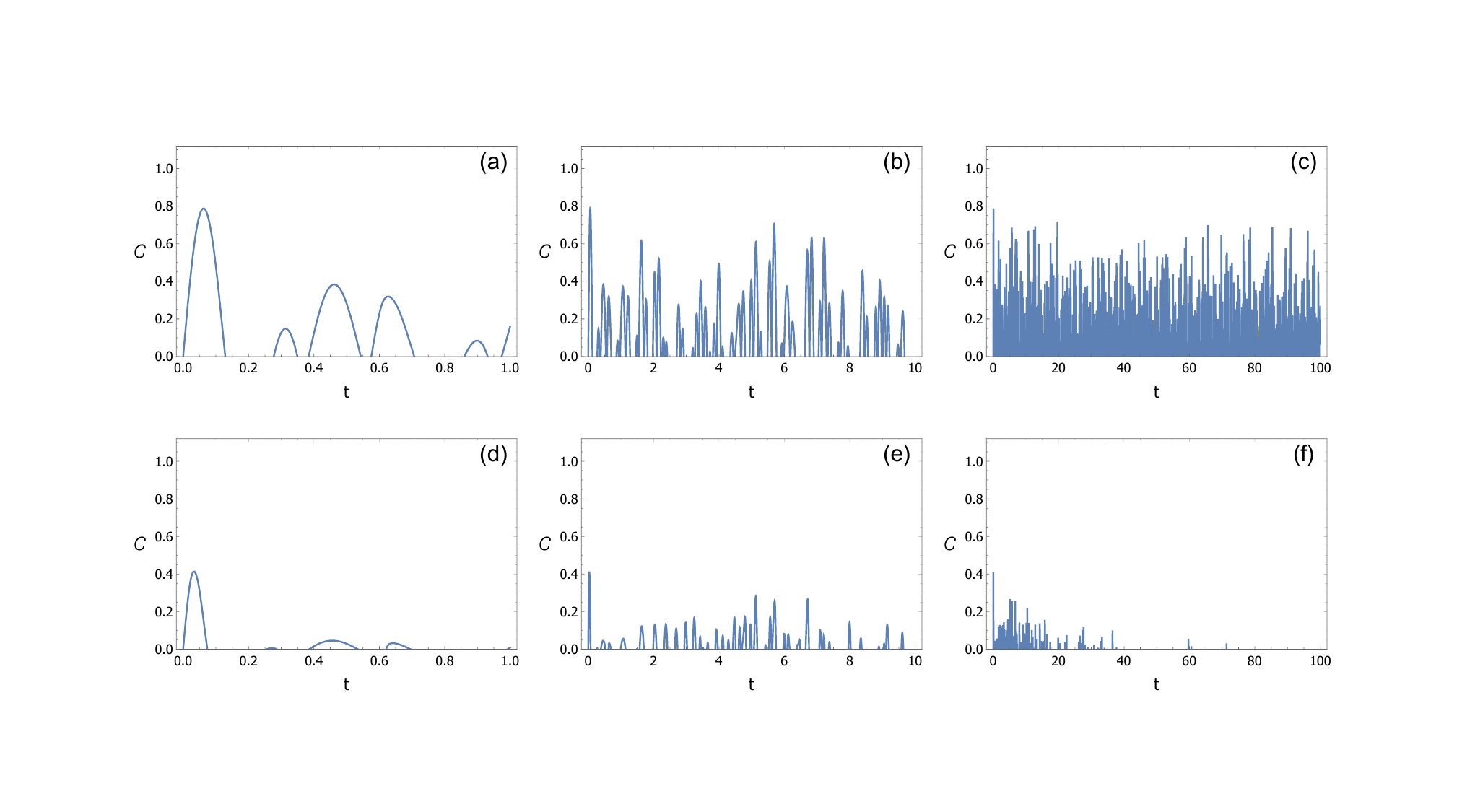}
	\caption{Concurrence of the two-qubit system as a function of time for $g = 0.5 \lambda$ and 
		for different time-scales. The mean
		photon number of the field is $\overline{n} = 1$ in (a), (b) and (c), 
		and $\overline{n} = 10$ in (d), (e) and (f).
		The initial state is the same as in Figure (1).} 
	\label{figure2}	
\end{figure}

\begin{figure}
	\centering
	\includegraphics[scale=0.8]{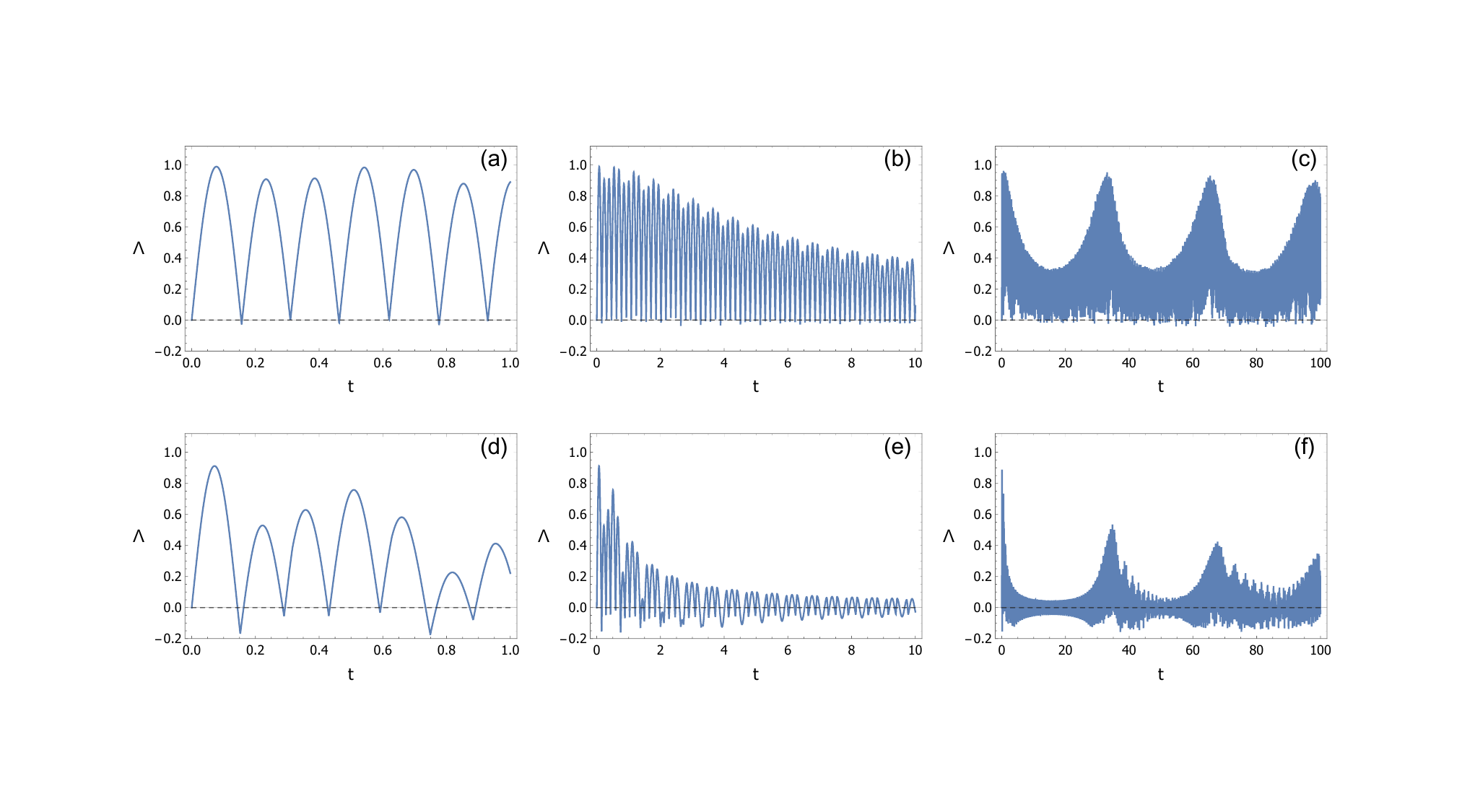}
	\caption{The $\Lambda$ function of the two-qubit system as a function of time for $g = 0.1 \lambda$ 
		and for different time-scales. The mean photon number of the field is 
		$\overline{n} = 1$ in (a), (b) and (c), and $\overline{n} = 10$ in (d), (e) and (f).
		The initial state is the same as in Figure (1).} 
	\label{figure3}	
\end{figure} 

\begin{figure}
	\centering
	\includegraphics[scale=0.8]{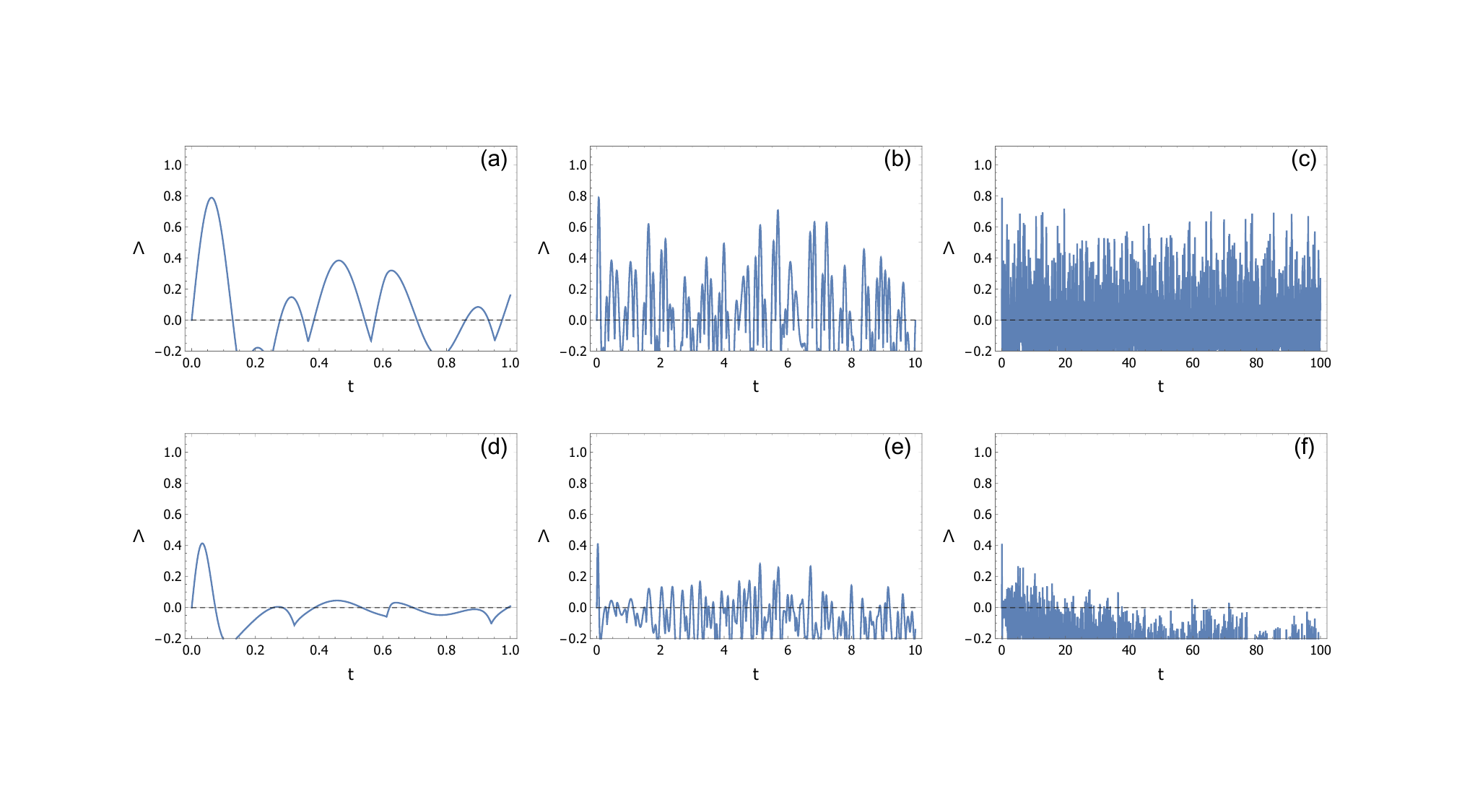}
	\caption{The $\Lambda$ function of the two-qubit system as a function of time for $g = 0.5 \lambda$ and 
		for different time-scales. The mean
		photon number of the field is $\overline{n} = 1$ in (a), (b) and (c), 
		and $\overline{n} = 10$ in (d), (e) and (f).
		The initial state is the same as in Figure (1).} 
	\label{figure4}	
\end{figure}

\begin{figure}
	\centering
	\includegraphics[scale=0.8]{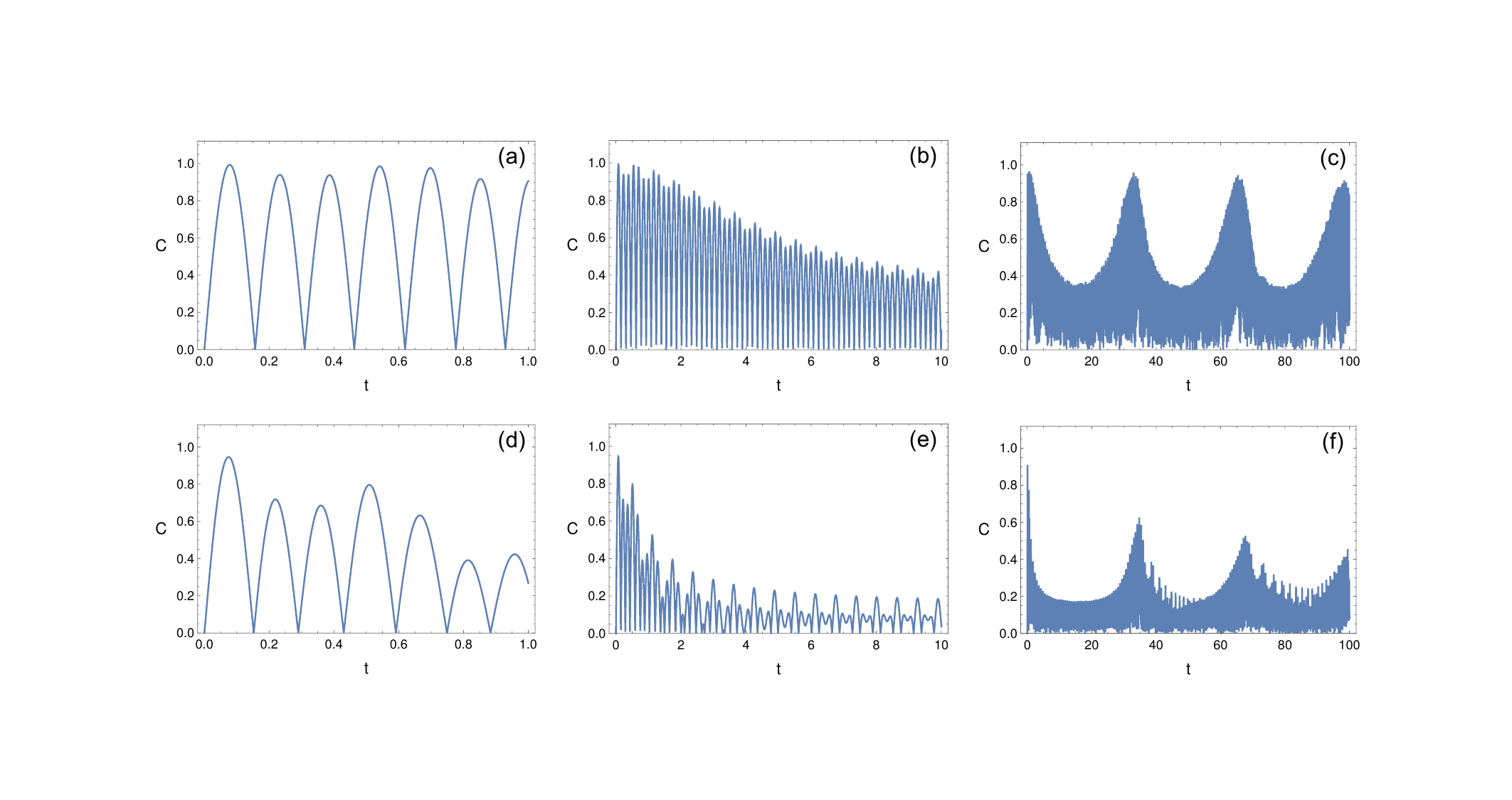}
	\caption{Quantum coherence of the two-qubit system as a function of time for $g = 0.1 \lambda$ and 
		for different time-scales. The mean
		photon number of the field is $\overline{n} = 1$ in (a), (b) and (c), 
		and $\overline{n} = 10$ in (d), (e) and (f).
		The initial state is the same as in Figure (1).} 
	\label{figure5}	
\end{figure} 

\begin{figure}
	\centering
	\includegraphics[scale=0.8]{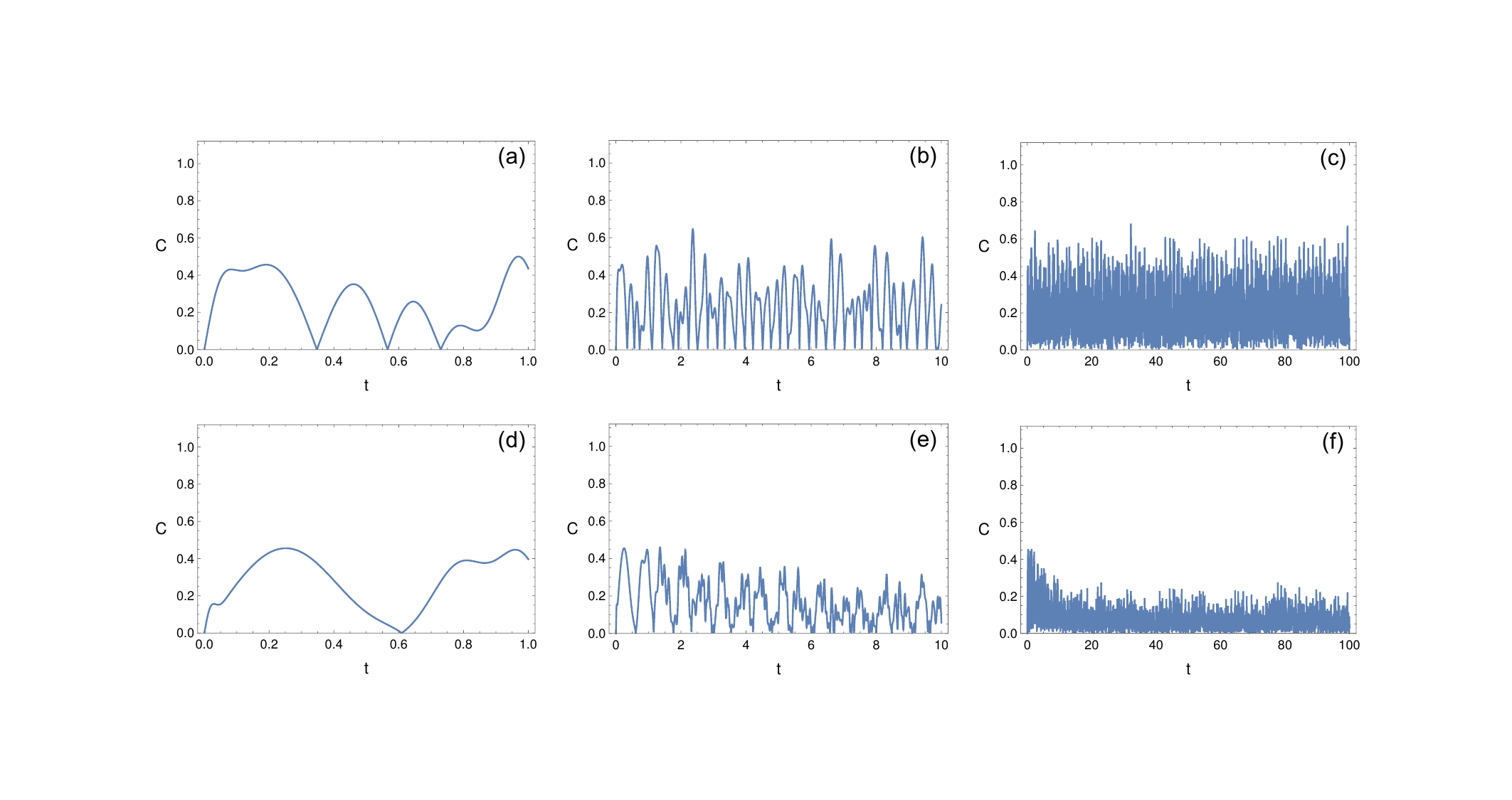}
	\caption{Quantum coherence of the two-qubit system as a function of time for $g = 0.5 \lambda$ and 
		for different time-scales. The mean
		photon number of the field is $\overline{n} = 1$ in (a), (b) and (c), 
		and $\overline{n} = 10$ in (d), (e) and (f).
		The initial state is the same as in Figure (1).} 
	\label{figure6}	
\end{figure}

\begin{figure}
	\centering
	\includegraphics[scale=0.8]{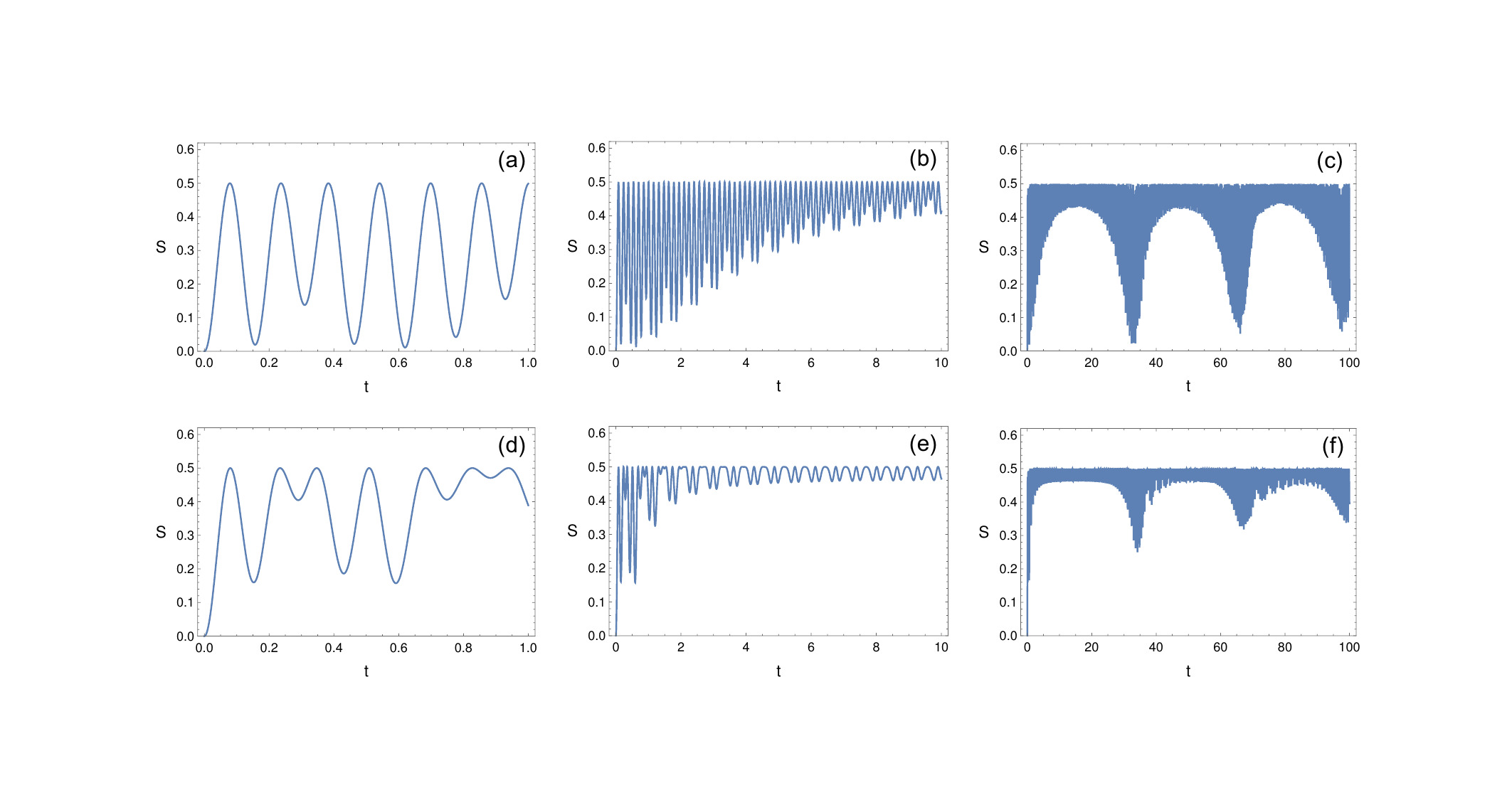}
	\caption{State purity of qubit 1 as a function of time for $g = 0.1 \lambda$ and for different time-scales. 
		The mean photon number of the field is $\overline{n} = 1$ in (a), (b) and (c), and $\overline{n} = 10$ 
		in (d), (e) and (f). The initial state is the same as in Figure (1).} 
	\label{figure7}	
\end{figure} 

\begin{figure}
	\centering
	\includegraphics[scale=0.8]{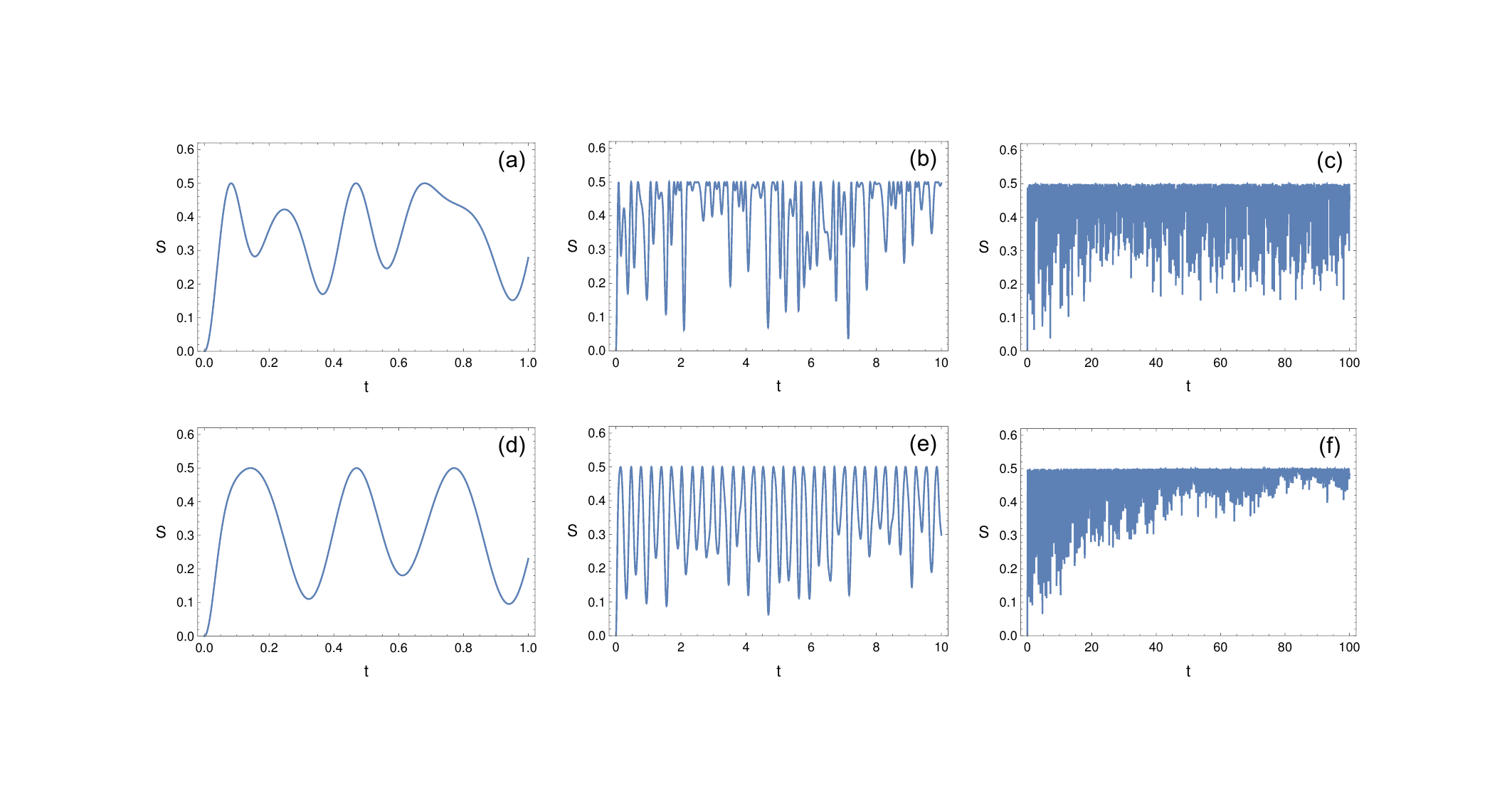}
	\caption{State purity of qubit 1 as a function of time for $g = 0.5 \lambda$ and for different time-scales. The mean
		photon number of the field is $\overline{n} = 1$ in (a), (b) and (c), and $\overline{n} = 10$ in (d), (e) and (f).
		The initial state is the same as in Figure (1).} 
	\label{figure8}	
\end{figure} 

\end{document}